\documentclass[aps,12pt,preprint]{revtex4}

\usepackage[dvips]{graphicx}
\usepackage{amsmath,amssymb,amsthm}

\newcommand{\eps}{\varepsilon}
\renewcommand{\bar}{\overline}

\DeclareMathOperator{\tr}{tr}
\DeclareMathOperator{\vol}{vol}

\begin{document}
\preprint{\hfill$\vcenter{\hbox{IUHET-515}}$}

\title{\vspace*{.75in}
Entanglement Entropy and Spatial Geometry}

\author{Micheal S. Berger}%
 \email{berger@indiana.edu}
\author{Roman V. Buniy}%
 \email{rbuniy@indiana.edu}
\affiliation{%
Physics Department, Indiana University, Bloomington, IN 47405, USA
}%

\date{\today}

\begin{abstract}
  The entanglement entropy in a quantum field theory between two
  regions of space has been shown in simple cases to be proportional
  to the volume of the hypersurface separating the regions. We prove
  that this is true for a free scalar field in an arbitrary geometry
  with purely spatial curvature and obtain a complete asymptotic
  expansion for the entropy.
\end{abstract}

\maketitle

\section{Introduction}

Modern developments in efforts to consistently combine gravity and
quantum mechanics have indicated that quantum field theory has too
many degrees of freedom. The entropy of a geometrical object seems to
depend on the area of a boundary surface. In the case of a black hole,
thermodynamic arguments suggest that its entropy is proportional to
the area of its event horizon. This expectation has been confirmed on
the microscopic side by calculations for some special cases of black
holes.

The separation of a system into two subsystems gives rise to the
notion of the entanglement entropy which quantifies the quantum
correlations between the two subsystems. One can take the subsystems
to be two regions of space separated by a boundary surface.  Early
calculations in quantum field theories indicated that the entanglement
entropy between the degrees of freedom in two regions separated by a
boundary surface is proportional to the area of the boundary
surface~\cite{Bombelli:1986rw,Srednicki:1993im,Callan:1994py,Fursaev:1994in}.
These developments have led to the intuition that the entanglement
entropy is dominated by degrees of freedom close to the boundary
surface, so it is natural to expect that the entanglement entropy is
proportional to the area. The connection between the entropy of a
geometric system, on the one hand, and the entanglement entropy
between the quantum field theory degrees of freedom between spatially
disconnected regions, on the other hand, is not clear. There have even
been suggestions that the black hole entropy is entirely entanglement
entropy~\cite{Kabat:1995eq}. This suggests the possibility that when
the boundary surface is taken to be an event horizon, the two types of
entropy are identical or at least related.  Therefore it becomes
important to examine the properties of entanglement entropy in more
general cases and attempt to understand its properties when the
entanglement involves regions separated by a horizon.

The entanglement entropy is ultraviolet divergent and must be
regularized. Presumably a more fundamental theory at the Planck scale
will provide the mechanism that eliminates the divergence. Since we
are not yet aware of how the underlying theory regularizes the
divergence, we are forced to do so by hand in calculations. The
regularization involves the Planck scale and the expectation that
entanglement entropy may be connected to the theory of quantum gravity
is suggested by simple dimensional analysis. If indeed the
entanglement entropy is proportional to the area of the boundary
surface in Planck units, then the connection to a holographic
principle similar to the one suggested by the AdS/CFT
correspondence~\cite{Maldacena:1997re,Witten:1998qj,Susskind:1998dq,Aharony:1999ti,Bousso:2002ju},
may apply to the entanglement entropy.

After the initial studies of the entanglement
entropy~\cite{Bombelli:1986rw,Srednicki:1993im,Callan:1994py,Fursaev:1994in},
most research in the subject has been devoted to CFTs. When a field
theory is conformal, there are additional tools to compute the
entanglement entropy, which, in fact, is proportional to the central
charge~\cite{Calabrese:2004eu}. Even when the CFT is modified by
adding mass
deformations~\cite{Calabrese:2004eu,Vidal:2002rm,Latorre:2003kg}, the
same property holds. Also, recent calculations of entanglement entropy
in this context suggest a holographic
interpretation~\cite{Ryu:2006bv,Ryu:2006ef}. The connection between
the entanglement entropy and holography has also been discussed in
more general contexts~\cite{Brustein:2005vx,Fursaev:2006ih}. Scaling
of the entaglement entropy with the area of the boundary have also
been verified by numerical
computations~\cite{Plenio:2004he,Casini:2003ix}.

In this paper, we investigate the entanglement entropy of a quantum
field theory in the case of an arbitrary boundary surface embedded in
a background with spatial curvature. Our results contain many
previously derived results as special cases, but are more general
because we consider arbitrary geometries. Ultimately one would want to
consider the even more general case involving spacetime curvature, so
the comparison with other forms of entropy can be made explicit.

\section{Entanglement Entropy\label{Entanglement-Entropy}}

In this section, we compute the entanglement entropy by using two main
tools, the replica method and the heat kernel method.

In the replica method (see, for example, Ref.~\cite{Callan:1994py}),
the entropy is expressed as the limit $k\to 1$ of an expression
involving the $k$th power of the (reduced) density matrix. Expressing
the density matrix as a path integral, we are lead to consider the
manifold which is the result of gluing $k$ copies of the original
manifold. This gives an expression of the entropy in terms of the path
integral over closed curves in the glued manifold, which is described
in the subsection~\ref{Replica-Method}.

For a free scalar field on an arbitrary base manifold, the resulting
path integral is expressed in terms of the spectral quantities of a
differential operator. It is convenient to study these quantities by
the heat kernel method. We use it in the subsection~\ref{Heat-Kernel}
to obtain an expression for the entropy in which the dependence on the
hypersurface is factored out.

This leads to several properties of the entropy which we derive in the
subsection~\ref{Several-Properties-of-Entropy}. In particular, we
compute the entropy for a hypersurface which is a direct product of
manifolds, prove the addititivity property of the entropy, and compute
the leading terms of asymptotic expansions of the entropy.

In the heat kernel method (see, for example,
Refs.~\cite{Gilkey:1995mj,Vassilevich:2003xt}), the spectral
information is obtained from an asymptotic expansion of the trace of
the heat kernel of the operator. All terms in the expansion are
determined by the geometry of the underlying manifold. Although they
can be computed in principle, the computations are quite complicated
in practice. The parameters in the resulting asymptotic expansion are
an ultraviolet cutoff scale and geometric scales associated with the
manifold. In the subsection~\ref{Asymptotics}, this leads to an
asymptotic expansion for the entropy, which involves geometric
quantities associated with the hypersurface. We show that the term
proportional to the volume of the manifold is absent and the leading
term is proportional to the volume of the hypersurface.

\subsection{Replica Method\label{Replica-Method}}

We consider a field theory on a generally curved space which is
divided by an arbitrary hypersurface into two parts. The quantum
fields in the two parts are entangled, and our goal is the calculation
of the entanglement entropy.  Let $M$ be an $n-1$ dimensional
Riemannian manifold without a boundary and let $\Sigma\subset M$ be a
closed hypersurface (a submanifold of codimension $1$). $\Sigma$
divides $M$ into two parts, the interior part $M'$ and the exterior
part $M''$. Let $\phi$ be a field on $M$, and $(\phi',\phi'')$ its
restrictions to $(M',M'')$, and let $\psi(\phi', \phi'')$ be a wave
function corresponding to the field having the value $(\phi', \phi'')$
on $(M', M'')$. The density matrix for $(\phi', \phi'')$ is
\begin{align}
  \rho(\phi'_1,\phi''_1,\phi'_2,\phi''_2) =\psi(\phi'_1,\phi''_1)
  \bar{\psi(\phi'_2,\phi''_2)},
\end{align}
and the reduced density matrix for $\phi'$ is obtained by tracing over
the degrees of freedom of the field on $M''$,
\begin{align}
  \rho'(\phi'_1,\phi'_2) =\int d\phi''\,
  \rho(\phi'_1,\phi'',\phi'_2,\phi'').
\end{align}
The reduced density matrix $\rho'$ represents the mixed state with the 
associated entropy
\begin{align}
  S'=-\tr{\biggl(\frac{\rho'}{\tr{\rho'}}
  \log{\frac{\rho'}{\tr{\rho'}}}\biggr)} =\lim_{k\to
  1}\biggl(1-\frac{\partial}{\partial k}\biggr)\log {\tr{\rho^{\prime
  k}}}.
\label{entropy}
\end{align}
The quantity $\tr{\rho'}=\int d\phi'\,\rho'(\phi',\phi')$ in the
denominator guarantees the correct normalization for the density
matrix. The second equality embodies the replica
method (see, for example, Ref.~\cite{Callan:1994py}).

In order to obtain the path integral representation for the $k$th
power of the density matrix we introduce an auxiliary field
$\varphi(\tau,x)$ defined on $N=\mathbb{R}\times M$ and which
satisfies the boundary condition $\varphi(0,x)=\phi_0(x)$. The parameter
$\tau$ represents Euclidean time.  Let $I(\varphi)$ be an action for
the field $\varphi$. The wave function is
$\psi(\phi)=Z(N,\phi_0,\phi)$, where
\begin{align}
  Z(N,\phi_0,\phi)=\int_{C(N,\phi_0,\phi)} d\varphi \,
  \exp{\bigl(-I(\varphi)\bigr)}
\end{align}
is a path integral over the space $C(N,\phi_0,\phi)$ of curves defined
on $N$ and which satisfy boundary conditions $\varphi(0,x)=\phi_0(x)$
and $\varphi(T,x)=\phi(x)$ for some $T\in\mathbb{R}$. Using
$T=-\infty$ for $\psi(\phi'_1,\phi'')$ and $T=\infty$ for
$\psi(\phi'_2,\phi'')$, we find 
\begin{align}
  \rho'(\phi'_1,\phi'_2) =Z(N,\phi'_1,\phi'_2).
\end{align}
The function $\varphi(\tau,x)$ has a discontinuity at $\tau=0$ since
$\varphi(0^-,x)=\phi'_1(x)$ and $\varphi(0^+,x)=\phi'_2(x)$. However,
$\varphi(\tau,x)$ is continuous on the manifold $\tilde{N}_1$, which
is defined as the manifold $N$ with the cut along $\{\tau=0\}\times
M'$.

\begin{figure}
  \centerline{\includegraphics[width=6in]{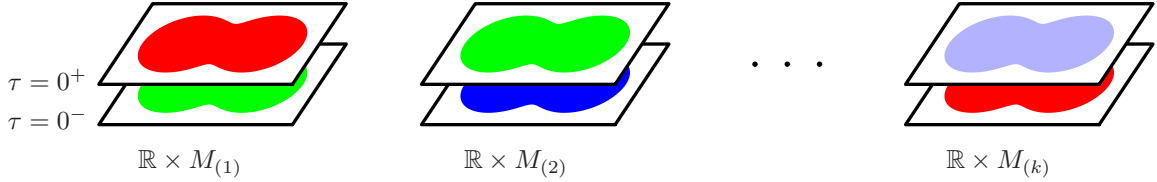}}
  \caption{The replica method involves cutting the original manifold
    $N$ along $\{\tau=0\}\times M'$ and gluing $k$ such cut copies of
    $N$ along $\{\tau=0\}\times M'$ to form the manifold $N_k$ with
    $k$ sheets. We identify $\{\tau_i=0^-\}\times M'_{(i)}$ with
    $\{\tau_{i+1}=0^+\}\times M'_{(i+1)}$ for $i=1,\ldots,k-1$, and
    $\{\tau_k=0^-\}\times M'_{(k)}$ with $\{\tau_1=0^+\}\times
    M'_{(1)}$. When $M=\mathbb{R}$ and $\Sigma$ is a point $P$, the
    construction gives the 2-dimensional cone manifold
    $C_k=\mathbb{R}^+\times S^1_k$, where $S^1_k$ is the unit circle
    $S^1$ which is parametrized by $0\le\theta\le 2\pi k$. The
    quantity $2\pi(1-k)$ is called the deficit angle. Note that $C_k$
    is the Riemann surface of the holomorphic function $z\mapsto
    z^k$.}
  \label{fig:geometry}
\end{figure}

The $k$th power of the density matrix is
\begin{align}
  \rho^{\prime k}(\phi'_1,\phi'_{k+1}) =\int d\phi'_2d\phi'_3\cdots d\phi'_k
  \, \rho^\prime (\phi'_1,\phi'_2)\rho^\prime (\phi'_2,\phi'_3) 
\cdots \rho^\prime (\phi'_k,\phi'_{k+1}).
\end{align}
Let $\mathbb{R}\times M_{(1)},\ldots,\mathbb{R}\times M_{(k)}$ be $k$
copies of $\mathbb{R}\times M$. By cutting every $\mathbb{R}\times
M_{(i)}$ along $\{\tau_i=0\}\times M'_{(i)}$ and gluing them in such a
way that $\{\tau_i=0^-\}\times M'_{(i)}$ is identified with
$\{\tau_{i+1}=0^+\}\times M'_{(i+1)}$ for $i=1,\ldots,k-1$, we obtain
the manifold $\tilde{N}_k$. See Fig.~\ref{fig:geometry}.
This gives
\begin{align}
  \rho^{\prime k}(\phi'_1,\phi'_{k+1})
  =Z(\tilde{N}_k,\phi'_1,\phi'_{k+1}).
\end{align}
Identifying $\{\tau_k=0^-\}\times M'_{(k)}$ with $\{\tau_1=0^+\}\times
M'_{(1)}$, we obtain the manifold $N_k$. This gives
\begin{align}
  \tr{\rho^{\prime k}}=Z(N_k)=\int_{C(N_k)} d\varphi \,
  \exp{\bigl(-I(\varphi)\bigr)},
\end{align}
which is a path integral over all closed curves in $N_k$. This
quantity gives the entanglement entropy for $\rho'$ via
Eq.~(\ref{entropy}). If instead we were to trace the density matrix
over the degrees of freedom in $M'$, we would obtained the reduced
density matrix $\rho''$ for $\phi''$.  It is easy to show that the
entanglement entropy for $\rho''$ is the same, $S'=S''$, and we denote
the common value by $S_\Sigma$ to emphasize its dependence on the
surface $\Sigma$.

\subsection{Heat Kernel\label{Heat-Kernel}}

To proceed with an explicit computation, we choose the free scalar
field with the action
\begin{align}
  I(\varphi)=2^{-1}\int_{N}\omega_N \varphi D_N\varphi,
\end{align}
where $D_N=\Delta_N+m^2$, $\Delta_N$ is the scalar Laplace operator
for $N$, $\omega_N$ is the volume form for $N$, and $m$ is the mass of
the field $\varphi$. Performing a Gaussian integral, we find
\begin{align}
  Z(N_k)=Z_0^k(\det{D_{N_k}})^{-1/2},
\end{align}
where $Z_0$ is a constant independent of $k$. Since $D_{N_k}$ is a
non-negative elliptic operator, we can define its determinant by
\begin{align}
  \log{\det{D_{N_k}}}-\log{\det{E_{N_k}}}=-\int_0^\infty
  dt\,t^{-1}\bigl(\tr{\exp{(-tD_{N_k})}}-\tr{\exp{(-tE_{N_k})}}\bigr),
  \label{det}
\end{align}
where $E_{N_k}$ is any other non-negative elliptic operator on
$N_k$. (To prove this equation, one writes the analogous equation
relating eigenvalues of $D_{N_k}$ and $E_{N_k}$.) The quantity
$\exp{(-tD_{N_k})}$ is called the heat kernel of the operator
$D_{N_k}$, and
\begin{align}
  K(t,D_{N_k})=\tr{\exp{(-tD_{N_k})}}
\end{align}
is its $L^2$ trace. We find
$K(t,D_{N_k})=\exp{(-tm^2)}K(t,\Delta_{N_k})$.

The integral over $t$ in Eq.~\eqref{det} diverges for small $t$. To
obtain a finite result, we replace the lower limit of integration over
$t$ by a regularization parameter $\lambda^2$ (an ultraviolet cutoff),
\begin{align}
  \int_{\lambda^2}^\infty
    dt\,t^{-1}K(t,D_{N_k})=\tr{\Gamma(0,\lambda^2 D_{N_k})}.
\end{align}
Here $\Gamma$ is the incomplete Gamma function which is given either
by the integral representation
\begin{align}
  \Gamma(\alpha,z)=\int_z^\infty du\, u^{\alpha-1}\exp{(-u)}
\end{align}
or by the series representation
\begin{align}
  \Gamma(\alpha,z)&=\Gamma(\alpha)
  -z^\alpha\sum_{j=0}^\infty\frac{(-z)^j}{(\alpha+j)j!},\quad
  \alpha\not=0,-1,-2,\ldots, \\
  \Gamma(-l,z)&=\frac{(-1)^l}{l!}(\psi(l+1)-\log{z})
  -z^{-l}\sum_{\begin{subarray}{c}j=0\\
  j\not=l\end{subarray}}^\infty\frac{(-z)^j}{(-l+j)j!},\quad
  l=0,1,2,\ldots,
\end{align}
where $\psi(l+1)=-\gamma+\sum_{j=1}^l j^{-1}$ and $\gamma$ is the
Euler constant. We will later need the incomplete Gamma function for
nonzero values of $\alpha$ as well.

We choose $E_{N_k}$ to be a unit operator times a constant with the
dimension of inverse length squared; this leads to vanishing of its
contribution to the entropy. Similarly, the contribution from the
constant $Z_0^k$ vanishes. The regularized entropy becomes
\begin{align}
  S_\Sigma(\lambda)=2^{-1}\lim_{k\to
  1}\biggl(1-\frac{\partial}{\partial k}\biggr)\tr{\Gamma(0,\lambda^2
  D_{N_k})}.
\end{align}

We can factor the dependence of $S_\Sigma(\lambda)$ on $\Sigma$
proceeding as follows. Locally, $N_k=C_k\times\Sigma$, where
$C_k=\mathbb{R}^+\times S^1_k$ is the 2-dimensional cone manifold, and
$S^1_k$ is the unit circle $S^1$ which is parametrized by
$0\le\theta\le 2\pi k$. The quantity $2\pi(1-k)$ is called the deficit
angle. Note that $C_k$ is the Riemann surface of the holomorphic
function $z\mapsto z^k$. Giving $N_k$ a product metric, we find
\begin{align}
  \Delta_{N_k}=\Delta_{C_k}\otimes
  1_\Sigma+1_{C_k}\otimes\Delta_\Sigma,
\end{align}
which gives
\begin{align}
  K(t,\Delta_{N_k})=K(t,\Delta_{C_k})K(t,\Delta_\Sigma).
\end{align}
This factorization reveals the special role played by the entropy for
a point $P$, when $M=\mathbb{R}$, $\Sigma=P$,
\begin{align}
  S_{P}(\lambda)=2^{-1}\int_{\lambda^2}^\infty dt\, t^{-1}\exp{(-t
  m^2)}C(t),
\end{align}
where
\begin{align}
  C(t)=\lim_{k\to 1}\biggl(1-\frac{\partial}{\partial
  k}\biggr)K(t,\Delta_{C_k}).\label{C}
\end{align}
A simple computation gives the expression
\begin{align}
  S_\Sigma(\lambda)=-\int_\lambda^\infty d\mu\, \frac{\partial
  S_{P}(\mu)}{\partial\mu}K(\mu^2,\Delta_\Sigma),
\end{align}
in which the dependence on $\Sigma$ is factored out. This equation
leads to several properties of the entropy, which we now derive.

\subsection{Several Properties of Entropy\label{Several-Properties-of-Entropy}}

\textbf{1.} Let $(r,\theta)$ be local polar coordinates for $C_k$, and
$\xi>0$. Under the scaling transformation $(t,r,\theta)\mapsto (\xi^2 t,\xi
r,\theta)$, we have $C_k\mapsto C_k$, $t\Delta_{C_k}\mapsto
t\Delta_{C_k}$, and so $K(t,\Delta_{C_k})\mapsto
K(t,\Delta_{C_k})$. This implies $C(t)=C=\text{const}$ and thus
\begin{align}
  S_{P}(\lambda)=2^{-1}C\Gamma(0,\lambda^2 m^2).\label{S0}
\end{align}
We will compute $C$ in the next subsection and the appendix.

\textbf{2.} Let $(\tau=x^1,x^2)$ be local coordinates for $C_k$, let
$(x^3,\ldots,x^{n})$ be local coordinates for $\Sigma$, and let
$\xi>0$. Under the transformation $\lambda\mapsto\xi\lambda$,
$x^j\mapsto \xi x^j$, $j=3,\ldots,n$, we have $\Delta_\Sigma\mapsto
\Delta_{\Sigma,\xi}=\xi^{-2}\Delta_\Sigma$ and
$S_\Sigma(\lambda)\mapsto S_{\Sigma,\xi}(\xi\lambda)$, where
\begin{align}
  S_{\Sigma,\xi}(\xi\lambda)=-\int_{\xi\lambda}^\infty d\mu\,
  \frac{\partial S_{P}(\mu)}{\partial\mu} \biggl(\frac{\partial
  S_{P}(\xi^{-1}\mu)}{\partial (\xi^{-1}\mu)}\biggr)^{-1}
  \frac{\partial S_\Sigma(\xi^{-1}\mu)}{\partial(\xi^{-1}\mu)}.
\end{align}
Using Eq.~\eqref{S0}, we find
\begin{align}
  S_{\Sigma,\xi}(\xi\lambda)=-\int_{\lambda}^\infty d\nu\,
  \exp{\bigl(-(\xi^2-1)\nu^2 m^2\bigr)} \frac{\partial
  S_\Sigma(\nu)}{\partial \nu}.
\end{align}
It follows that $\lim_{\xi\to 0}(\partial
S_{\Sigma,\xi}(\xi\lambda)/\partial\xi)=0$. Since
$\vol{(\Sigma)}\mapsto\xi^{n-2}\vol{(\Sigma)}$, this implies
\begin{align}
  S_\Sigma(\lambda)\sim C'\lambda^{2-n}\vol{(\Sigma)}, \quad
  \lambda\to 0,\label{SSigmaasympt}
\end{align}
where $C'=\text{const}$.

\textbf{3.} Since $0$ is the smallest eigenvalue of $\Delta_\Sigma$,
we have $K(t,\Delta_\Sigma)\sim 1$, $t\to\infty$. This gives
\begin{align}
  S_\Sigma(\lambda)\sim 2^{-1}C\Gamma(0,\lambda^2 m^2), \quad
  \lambda\to\infty.
\end{align}
Interestingly, this coincides with the expression in Eq.~\eqref{S0}
for $S_{P}(\lambda)$ for arbitrary $\lambda$.  This can be
understood as the result of the physical scales of $\Sigma$ becoming
irrelevant as the cutoff $\lambda$ tends to infinity.

\begin{figure}
  \centerline{ \includegraphics[width=1.5in]{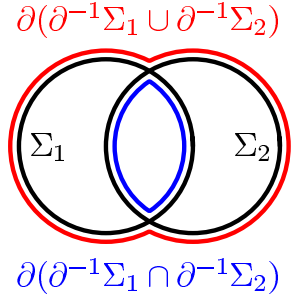} }
  \caption{The hypersurfaces used in the formulation of the additivity
  property.}
  \label{fig:additivity}
\end{figure}

\textbf{4.} Let $\Sigma_1$ and $\Sigma_2$ be closed hypersurfaces in
$M$. Let $\partial$ and $\partial^{-1}$ be operators defined by
$\partial M'_i=\Sigma_i$ and $\partial^{-1}\Sigma_i=M'_i$, where
$M'_i$ is a part of $M$ inside $\Sigma_i$ for $i=1,2$. Being an
integral over $\Sigma$, the quantity $K(t,\Delta_\Sigma)$ is linear in
$\Sigma$. It follows that
\begin{align}
  K(t,\Delta_{\Sigma_1})+K(t,\Delta_{\Sigma_2})
  =K(t,\Delta_{\partial(\partial^{-1}\Sigma_1\cup\partial^{-1}\Sigma_2)})
  +K(t,\Delta_{\partial(\partial^{-1}\Sigma_1\cap\partial^{-1}\Sigma_2)}),
\end{align}
and thus the entropy satisfies the additivity property
\begin{align}
  S_{\Sigma_1}(\lambda)+S_{\Sigma_2}(\lambda)
  =S_{\partial(\partial^{-1}\Sigma_1\cup\partial^{-1}\Sigma_2)}(\lambda)
  +S_{\partial(\partial^{-1}\Sigma_1\cap\partial^{-1}\Sigma_2)}(\lambda).
  \label{additivity}
\end{align}
See Fig.~\ref{fig:additivity}.
For an arbitrary system, the entanglement entropy satisfies the strong
subadditivity property, which requires `$\ge$' instead of `$=$' in
Eq.~\eqref{additivity}.

\subsection{Asymptotics\label{Asymptotics}}

We now derive the asymptotic expansion of $S_\Sigma(\lambda)$ for
$\lambda\to 0$. It is clear that this requires knowledge of the
asymptotic behavior of $K(t,\Delta_\Sigma)$ for $t\to 0$. For an
$n$-dimensional manifold $L$, such an asymptotic is given by
\begin{align}
  K(t,\Delta_L)\sim \sum_{l=0}^\infty t^{(l-n)/2} a_l(\Delta_L), \quad
  t\to 0,
\end{align} 
where
\begin{align}
  a_l(\Delta_L)=\int_L \omega_L a_l(x_L,\Delta_L),
\end{align} 
and $a_l(x_L,\Delta_L)$ are the heat kernel coefficients for
$\Delta_L$. The above factorization of $K(t,\Delta_{N_k})$ leads to
\begin{align}
  a_l(x_{N_k},\Delta_{N_k})=\sum_{j=0}^l a_j(x_{C_k},
  \Delta_{C_k})a_{l-j}(x_\Sigma,\Delta_\Sigma).
\end{align}

The coefficients $a_l(x_L,\Delta_L)$ are completely determined by the
geometry of $L$. For a manifold without boundary,
$a_l(x_L,\Delta_L)=0$ for odd $l$. All coefficients
$a_l(x_L,\Delta_L)$ are polynomials in the covariant derivatives of
the Riemann tensor $(R_L)_{abcd}$, the Ricci tensor $(R_L)_{ab}$, and
the scalar curvature $R_L$ of $L$. Explicit expressions for several
first coefficients are available in the literature (see, for example,
Ref.~\cite{Gilkey:1995mj}). For example,
\begin{align}
  a_0(x_L,\Delta_L)&=(4\pi)^{-n/2},\label{a0}\\
  a_2(x_L,\Delta_L)&=(4\pi)^{-n/2}6^{-1}R_L,\label{a2}\\
  a_4(x_L,\Delta_L)&=(4\pi)^{-n/2}360^{-1}\Bigl(-12\Delta_L R_L
  +5R_L^2 -2\sum_{a,b}(R_L)_{ab}(R_L)_{ab}\nonumber\\
  &+2\sum_{a,b,c,d}(R_L)_{abcd}(R_L)_{abcd}\Bigr).
\end{align}

It has been shown~\cite{Cheeger} that the only nonzero heat kernel
coefficients for $C_k$ are
\begin{align}
  a_0(x_{C_k},\Delta_{C_k})&=(4\pi)^{-1},\\
  a_2(x_{C_k},\Delta_{C_k})&=(4\pi)^{-1}6^{-1}4\pi(1-k)\delta_{C_k},
\end{align}
where $\delta_{C_k}$ is the delta function at the origin of
$C_k$. This gives $C=6^{-1}$. (In the appendix, we derive this
result.)  The regularized entropy becomes
\begin{align}
  S_{P}(\lambda)&= 12^{-1}\Gamma(0,\lambda^2 m^2),\\
  S_\Sigma(\lambda)&= 12^{-1}\tr{\Gamma(0,\lambda^2 D_\Sigma)}.
\end{align}
In terms of the integrated heat kernel coefficients of $\Sigma$, we
find
\begin{align}
  S_\Sigma(\lambda)\sim 12^{-1}\sum_{l=0}^\infty
  m^{n-l-2}\Gamma\bigl((2+l-n)/2,(\lambda m)^2\bigr)
  a_l(\Delta_\Sigma), \quad \lambda\to 0.
\end{align}
This asymptotic expansion is our main result. For $\lambda\to 0$,
$S_\Sigma(\lambda)$ depends only on the spectral properties of the
operator $\lambda^2 D_\Sigma$. Equivalently, the entropy depends only
on parameters $m$, $\lambda$, and on geometric invariants associated
with $\Sigma$. The asymptotic expansion for $S_\Sigma(\lambda)$
involves $\log{\lambda m}$ and the powers of $\lambda m$. The leading
term in the entropy is
\begin{align}
  S_{P}(\lambda)&\sim 12^{-1}\bigl(-2\log{\lambda m}-\gamma\bigr),
  \quad \lambda\to 0,\label{entropy-P-asymptotic}\\
  S_\Sigma(\lambda)&\sim
  12^{-1}(n/2-1)^{-1}\lambda^{2-n}(4\pi)^{1-n/2}\vol{(\Sigma)}, \quad
  \lambda\to 0.\label{entropy-Sigma-asymptotic}
\end{align}
The term of order $\lambda^{-n}\vol{(N)}$ in $S_\Sigma(\lambda)$ is
absent; it would be the extensive contribution to the entropy.

We remark on the case $n=2$. (See also Ref.~\cite{Ryu:2006ef} for a
similar discussion.) The entanglement entropy for a critical
2-dimensional CFT with the central charge $c$ is asymptotically
$S\sim(c/3)\log{(\ell/\lambda)}$, where $\ell$ is the size of the
system~\cite{Calabrese:2004eu}. For a massive theory with the
correlation length $\xi$, the entropy becomes
$S\sim(c/6)\nu\log{(\xi/\lambda)}$ for $\ell\gg\xi$, where $\nu$ is the
number of components of (zero dimensional) $\Sigma$. Setting $c=1$,
$\nu=1$, $\xi\sim m^{-1}$, we recover
Eq.~\eqref{entropy-P-asymptotic}.

\section{Conclusion}

We have calculated the asymptotic expansion of the entanglement
entropy for a free scalar field in arbitrary background geometry. The
expansion parameter is the ultraviolet cutoff $\lambda$ which is
needed to regularize the entropy. We have found that the entropy
depends only on geometric invariants associated with the boundary
surface $\Sigma$. The extensive contribution to the entropy, the term
of order $\lambda^{-n}\vol{(N)}$, is absent. The leading term is
proportional to $\lambda^{2-n}\vol{(\Sigma)}$.

We have considered a situation with spatial curvature only and with
time included in only a trivial way as a product. Our calculation does
not utilize a spacetime that is a solution of Einstein's
equations. Further research may involve extending this calculation to
cases involving spacetime curvature. Another interesting direction to
pursue is to include interactions since this, at least intuitively,
can potentially change the area dependence of the entropy. The studies
of interactions have been mostly limited to CFTs, but their role in
QFT remains largely unexplored. These cases involve gravity more
explicitly and may make a connection between holographic entropy and
entanglement entropy more obvious.

\begin{acknowledgments}

We thank Tom Kephart, Juan Maldacena, and Tadashi Takayanagi for
discussions.  This work was supported in part by the U.S.  Department
of Energy under Grant No.~DE-FG02-91ER40661.

\end{acknowledgments}

\appendix*

\section{}

Here we compute the heat kernel coefficients for the cone $C_k$. Let
$\xi>0$. Under the transformation $(t,r,\theta)\mapsto (\xi^2 t,\xi
r,\theta)$, we have $K(t,\Delta_{C_k})\mapsto K(t,\Delta_{C_k})$,
$a_l(\Delta_{C_k})\mapsto\xi^{2-l}a_l(\Delta_{C_k})$. Since $C_k$ does
not have a length scale associated with it, this implies
$a_0(\Delta_{C_k})=\infty$, $a_2(\Delta_{C_k})=\text{const}$,
$a_l(\Delta_{C_k})=0$, $l\ge 4$. $a_0(x_{C_k},\Delta_{C_k})$ is given
by Eq.~\eqref{a0}, and to compute $a_2(x_{C_k},\Delta_{C_k})$ from
Eq.~\eqref{a2}, we need to know the scalar curvature of $C_k$, with
computation of which we now proceed.

$C_k$ is singular at $r=0$ if $k\not=1$. We consider it as a limit
$C_k=\lim_{\eps\to 0}C_{k,\eps}$, where $C_{k,\eps}$ is a regular
manifold. On $C_{k,\eps}$ we take an orthonormal frame
$(\omega^1,\omega^2)=(fdr,rd\theta)$, where the regularization
function $f(k,r,\eps)$ is an arbitrary smooth function satisfying
conditions $\lim_{r\to 0}f=k$, $\lim_{\eps\to 0}f=1$. An example of
such a function is $f=\bigl(k^2+(1-k^2)(qr)^\eps\bigr)^{1/2}$, where
$\eps\ge 0$ and $q>0$ is an arbitrary constant with the dimension of
inverse length. In what follows, we proceed with arbitrary $f$
satisfying the above conditions.

Let $\omega$ and $\Omega$ be the $2\times 2$ antisymmetric matrices of
connection and curvature forms. Cartan's equations for $C_{k,\eps}$,
\begin{align}
  {\omega^1}_2\wedge r d\theta&=0,\\ 
  dr\wedge d\theta+{\omega^2}_1\wedge fdr&=0,\\
  d{\omega^1}_2&={\Omega^1}_2,
\end{align}
have the solution 
\begin{align}
  {\omega^1}_2&=-f^{-1}d\theta,\\ 
  {\Omega^1}_2&=rfg dr\wedge d\theta,
\end{align}
where $g=r^{-1}f^{-3}(\partial f/\partial r)$. The nonzero components
of the Ricci tensor are
$(R_{C_{k,\eps}})_{11}=(R_{C_{k,\eps}})_{22}=g$, and the scalar
curvature is $R_{C_{k,\eps}}=2g$.

To obtain non-regularized quantities, we consider the limit $\eps\to
0$. Using $\lim_{r\to 0}f=k$, $\lim_{\eps\to 0}f=1$,
$\omega_{C_{k,\eps}}=rfdr\wedge d\theta$, for an arbitrary function
$h(r)$ satisfying $h(\infty)=0$, we find
\begin{align}
  \lim_{\eps\to 0}\int_{C_{k,\eps}}\omega_{C_{k,\eps}}gh =2\pi k
  \lim_{\eps\to 0}\biggl(-f^{-1}h\bigl\vert_{r=0}^{r=\infty}
  +\int_0^\infty dr\,f^{-1}(\partial h/\partial r)\biggr)
  =2\pi(1-k)h(0).
\end{align}
This implies $\lim_{\eps\to 0}g=2\pi(1-k)\delta_{C_k}$, where
$\delta_{C_k}$ is the delta function at the origin of $C_k$. Thus,
$R_{C_k}=4\pi(1-k)\delta_{C_k}$, the only nonzero heat kernel
coefficients for $C_k$ are
\begin{align}
  a_0(x_{C_k},\Delta_{C_k})&=(4\pi)^{-1},\\
  a_2(x_{C_k},\Delta_{C_k})&=(4\pi)^{-1}6^{-1}4\pi(1-k)\delta_{C_k},
\end{align}
so that we may identify $C=6^{-1}$. Alternatively, since $C(t)$ in
Eq.~\eqref{C} is a constant, we can compute it by taking the limit
$t\to 0$ in the expansion
\begin{align}
  C(t)\sim\lim_{k\to 1}\biggl(1-\frac{\partial}{\partial k}\biggr)
  \sum_{l=0}^\infty t^{(l-2)/2} a_l(\Delta_{C_k}), \quad t\to 0.
\end{align}
Since the coefficients $a_0(x_{C_k},\Delta_{C_k})$,
$a_l(x_{C_k},\Delta_{C_k})$, $l\ge 4$ do not contribute to $C$, we
find $C=6^{-1}$.


\begin{thebibliography}{99}





\bibitem{Bombelli:1986rw}
  L.~Bombelli, R.~K.~Koul, J.~H.~Lee and R.~D.~Sorkin,
  Phys.\ Rev.\  D {\bf 34}, 373 (1986).

\bibitem{Srednicki:1993im}
  M.~Srednicki,
  Phys.\ Rev.\ Lett.\  {\bf 71}, 666 (1993)
  [arXiv:hep-th/9303048].

\bibitem{Callan:1994py}
  C.~G.~Callan and F.~Wilczek,
  Phys.\ Lett.\  B {\bf 333}, 55 (1994)
  [arXiv:hep-th/9401072];
  C.~Holzhey, F.~Larsen and F.~Wilczek,
  Nucl.\ Phys.\  B {\bf 424}, 443 (1994)
  [arXiv:hep-th/9403108].

\bibitem{Fursaev:1994in}
  D.~V.~Fursaev,
  Phys.\ Lett.\  B {\bf 334}, 53 (1994)
  [arXiv:hep-th/9405143];
  D.~V.~Fursaev and S.~N.~Solodukhin,
  Phys.\ Rev.\  D {\bf 52}, 2133 (1995)
  [arXiv:hep-th/9501127].

\bibitem{Kabat:1995eq}
  D.~Kabat,
  Nucl.\ Phys.\  B {\bf 453}, 281 (1995)
  [arXiv:hep-th/9503016].

\bibitem{Maldacena:1997re}
  J.~M.~Maldacena,
  Adv.\ Theor.\ Math.\ Phys.\  {\bf 2}, 231 (1998)
  [Int.\ J.\ Theor.\ Phys.\  {\bf 38}, 1113 (1999)]
  [arXiv:hep-th/9711200].

\bibitem{Witten:1998qj}
  E.~Witten,
  Adv.\ Theor.\ Math.\ Phys.\  {\bf 2}, 253 (1998)
  [arXiv:hep-th/9802150].

\bibitem{Susskind:1998dq}
  L.~Susskind and E.~Witten,
  arXiv:hep-th/9805114.

\bibitem{Aharony:1999ti}
  O.~Aharony, S.~S.~Gubser, J.~M.~Maldacena, H.~Ooguri and Y.~Oz,
  Phys.\ Rept.\  {\bf 323}, 183 (2000)
  [arXiv:hep-th/9905111].

\bibitem{Bousso:2002ju}
  R.~Bousso,
  Rev.\ Mod.\ Phys.\  {\bf 74}, 825 (2002)
  [arXiv:hep-th/0203101].

\bibitem{Calabrese:2004eu}
  P.~Calabrese and J.~L.~Cardy,
  J.\ Stat.\ Mech.\  {\bf 0406}, P002 (2004)
  [arXiv:hep-th/0405152].

\bibitem{Vidal:2002rm}
  G.~Vidal, J.~I.~Latorre, E.~Rico and A.~Kitaev,
  Phys.\ Rev.\ Lett.\  {\bf 90}, 227902 (2003)
  [arXiv:quant-ph/0211074].

\bibitem{Latorre:2003kg}
  J.~I.~Latorre, E.~Rico and G.~Vidal,
  Quant.\ Inf.\ Comput.\  {\bf 4}, 48 (2004)
  [arXiv:quant-ph/0304098].

\bibitem{Ryu:2006bv}
  S.~Ryu and T.~Takayanagi,
  Phys.\ Rev.\ Lett.\  {\bf 96}, 181602 (2006)
  [arXiv:hep-th/0603001].

\bibitem{Ryu:2006ef}
  S.~Ryu and T.~Takayanagi,
  JHEP {\bf 0608}, 045 (2006)
  [arXiv:hep-th/0605073].

\bibitem{Brustein:2005vx}
  R.~Brustein, M.~B.~Einhorn and A.~Yarom,
  JHEP {\bf 0601}, 098 (2006)
  [arXiv:hep-th/0508217].

\bibitem{Fursaev:2006ih}
  D.~V.~Fursaev,
  JHEP {\bf 0609}, 018 (2006)
  [arXiv:hep-th/0606184].

\bibitem{Plenio:2004he}
  M.~B.~Plenio, J.~Eisert, J.~Dreissig and M.~Cramer,
  Phys.\ Rev.\ Lett.\  {\bf 94}, 060503 (2005)
  [arXiv:quant-ph/0405142].

\bibitem{Casini:2003ix}
  H.~Casini,
  Class.\ Quant.\ Grav.\  {\bf 21}, 2351 (2004)
  [arXiv:hep-th/0312238].

\bibitem{Gilkey:1995mj} P.~B.~Gilkey, {\it Invariance Theory, the Heat
  Equation, and the Atiyah-Singer Index Theorem}, 2nd ed., CRC Press,
  Boca Raton (1994).

\bibitem{Vassilevich:2003xt}
  D.~V.~Vassilevich,
  Phys.\ Rept.\  {\bf 388}, 279 (2003)
  [arXiv:hep-th/0306138].

\bibitem{Cheeger}
  J.~Cheeger,
  J.\ Diff.\ Geom. {\bf 18}, 575 (1983).





\end{thebibliography}
\end{document}